\documentclass[preprint,prl,showpacs,byrevtex,superscriptaddress]{revtex4-1}
\usepackage{graphicx}
\usepackage{epsfig}

\begin{document}

\title{Random Walk by Majority Rule and L{\'e}vy walk}
\author{Hyungseok Chad Moon}
\affiliation{Department of Physics and Astronomy, Seoul National University, Seoul 08826, Korea}
\author{Kyungsun Moon}
\affiliation{Department of Physics, Yonsei University, Seoul 03722, Korea, E-mail: kmoon@yonsei.ac.kr}

\date{\today}

\begin{abstract}
We have studied a random walk model based on majority rule. At a given instant, the moving direction of a cargo is determined by motor coordination mediated by a {\em tug-of-war} mechanism between two kinds of competing motor proteins. We have demonstrated that the probability distribution $P(t)$ for unidirectional run time $t$ of a cargo can be remarkably described by L{\'e}vy walk for $t<\gamma_u^{-1}$ as $P(t)\propto t^{-3/2} e^{-\gamma_u t}$ with $\gamma_u$ being the unbinding rate of a motor protein from microtubule. The mean squared displacement of a cargo changes from super-diffusive behavior $\langle X^2\rangle\propto t^2$ for $t<\gamma_u^{-1}$ to normal diffusion $\langle X^2\rangle\propto t$ for $t>\gamma_u^{-1}$.
By considering the correlation effect in binding of a motor protein to microtubule, we have shown that L{\'e}vy walk behavior of $P(t)\propto t^{-{3/2}}$ persists robustly against correlations only adding an effective cutoff time $\gamma_b/\gamma_c^2$ with $\gamma_c$ representing the amount of correlations.

\end{abstract}
\maketitle

Motions of an immotile minute particle suspended in liquid demonstrate Brownian dynamics, which reflects on the random recoils imparted by constituent molecules.\cite{Brown} Random walk is quite useful in describing the stochastic dynamics of the particle, which can naturally explain the normal diffusive behavior of the particle.\cite{Einstein} In contrast, the motility of living species is much more complicated due to their intrinsic complexity and interactions with their environment.\cite{Mendez,Viswanathan} There are growing number of instances, which demonstrate L{\'e}vy walk-like behavior for their trajectories covering many different length scales from soil amoeba to human walk pattern. In L{\'e}vy walk statistics, the run time distribution is given by power-law $P(t)\propto t^{-\alpha}$ and hence has a heavy-tail.\cite{Klafter,Rhee,Zaburdaev,Jeon}

\begin{figure}
\epsfxsize=7.in \epsfysize=4.in \epsffile{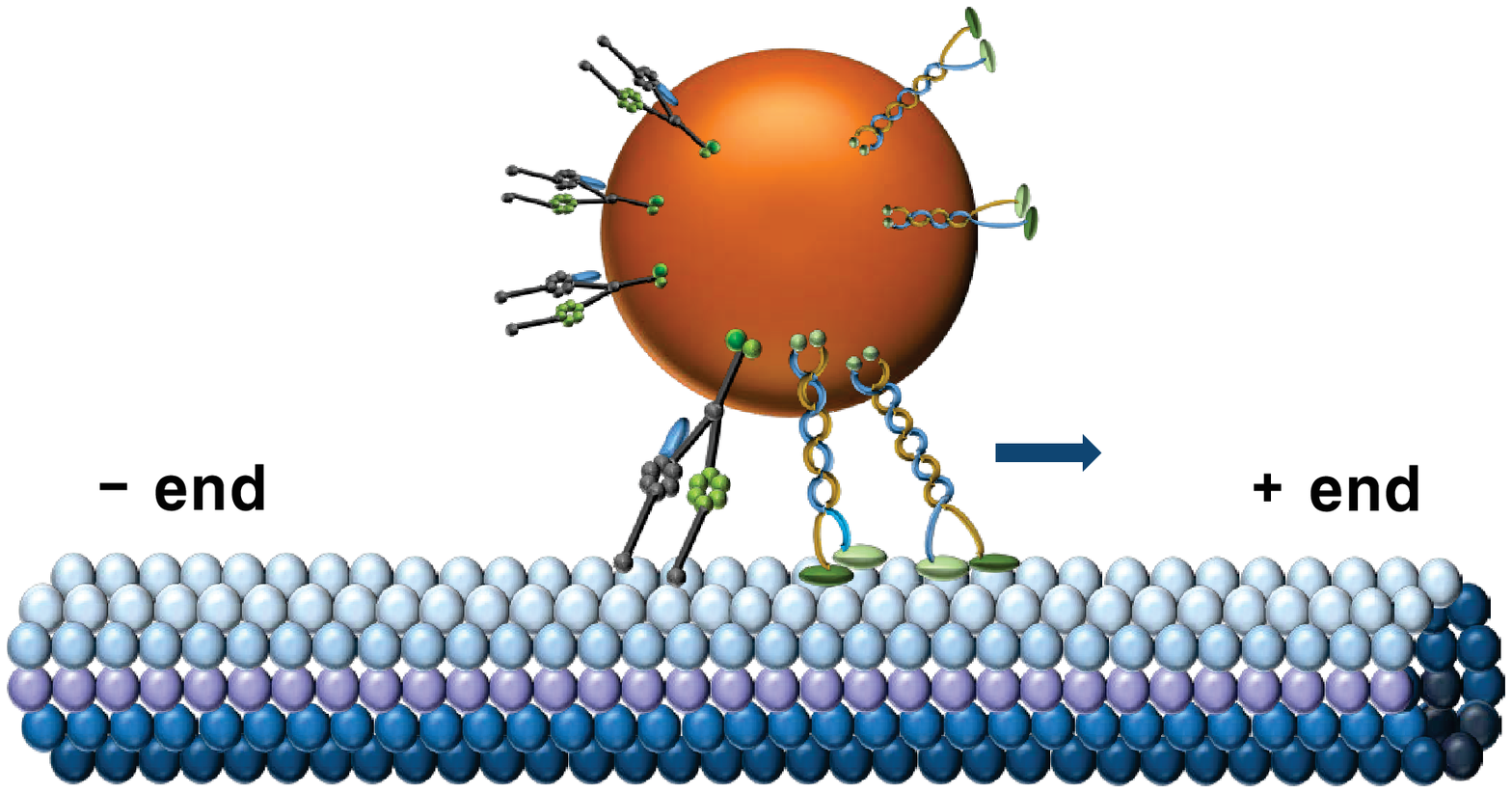}
\caption{(Color online) Two representative motor proteins: Kinesin and dynein. Four kinesins and four dyneins are attached to the cargo. Among them, two kinesins and one dynein are bound to the microtubule. By motor coordination via a {\em tug-of-war} mechanism, the cargo moves to the plus end of microtubule.}
\label{Motorprotein}
\end{figure}

At the cellular level, it has been known for long time that during saltatory intracellular particle movements, the particles may suddenly move many microns in seconds after short random irregular motions. The saltatory motion is much faster than Brownian motion and shown to be present in a variety of cells.\cite{Rebhun} For the intracellular transport, motor proteins actively participate in the transport of a variety of intra-cellular cargos, including vesicles, organelles, protein complexes along microtubules with definite polarity.\cite{Block,Gross,Kunwar}
Kinesins and dyneins are two representative motor proteins that are actively involved in microtubule transport. In general, kinesins with N-terminal motor domains move their cargo towards the plus ends of microtubules, while dyneins move their cargo towards the minus ends of microtubules as shown in Fig.~\ref{Motorprotein}. Hence kinesins or dyneins alone will be responsible only for the unidirectional cargo transport.

Recent advances in fluorescence imaging technique with high temporal resolution have revealed that the bi-directional motions of a cargo are frequently observed in microtubule transport.\cite{Chen,Park,Svoboda,Kojima,Visscher,Lipowsky,Granick,Wieschaus,Mallik} The bi-directional transport of a cargo has been generally attributed to the simultaneous presence of both kinesins and dyneins attached to the cargo.\cite{Lipowsky,Hancock} The trajectories of a cargo in intracellular transport along microtubule have quite often exhibited a super-diffusive behavior. In order for motor proteins to actively participate in cargo transport, they should bind to the microtubule.\cite{Svoboda,Lipowsky}
Since the binding process of a motor protein to microtubule is presumed to be stochastic, it is quite unlikely to expect a super-diffusive behavior followed by L{\'e}vy walk with long time correlation. Along with the mysterious emergence, the nature of L{\'e}vy walk observed in the experiment has posed very fundamental and important open question to be answered.

In our work, we have studied the cargo transport along microtubule by motor coordination mediated by a {\em tug-of-war} mechanism. The probability distribution $P(t)$ for the unidirectional transport of a cargo demonstrates L{\'e}vy walk behavior $P(t)\propto t^{-{3/2}} e^{-\gamma_u t}$ with a finite cutoff time $\gamma_u^{-1}$. The mean squared displacement of a cargo changes from super-diffusive behavior $\langle X^2\rangle\propto t^2$ for $t<\gamma_u^{-1}$ to normal diffusion $\langle X^2\rangle\propto t$ for $t>\gamma_u^{-1}$. For the first time, we have also analyzed the effect of correlations in binding process of a motor protein to microtubule. One can expect that the power-law exponent of L{\'e}vy walk may change due to correlations. Contrary to our intuition, we have noticed that the L{\'e}vy walk behavior of $P(t)\propto t^{-3/2}$ remains to hold against correlations only adding an effective cutoff time $\gamma_b/\gamma_c^2$ with $\gamma_c$ representing the amount of correlations.

We will explain our model for a microtubule-based transport of a cargo, to which two kinds of motor proteins are attached. At a given instant, a motor protein can bind stochastically to the microtubule at the rate of $\gamma_b$. For the sake of simplicity, we will assume that all the physical properties of two kinds of motor proteins are identical except that one kind of motor protein, which we will call `plus motor', moves to the plus-end of microtubule and the other one, which we will call `minus motor', moves to the minus-end of microtubule. Once a motor protein binds to the microtubule, it begins to actively participate in an attempt to move the cargo towards its own moving direction.
At a given instant, the moving direction of a cargo is determined by the majority between two kinds of active motor proteins, which have been bound to the microtubule until then. An active motor protein may unbind from the microtubule at the unbinding rate of $\gamma_{u}$, which is typically much slower than $\gamma_b$. Then it stops to participate in moving the cargo. In biological community, this class of models is generally dubbed as a {\em tug-of-war} model for bi-directional transport of a cargo by motor proteins.\cite{Lipowsky}

We will assume that the cargo was initially at rest until a single plus motor binds to the microtubule at $t=0$. It moves the cargo to the plus-end of microtubule at velocity $v$ during $\Delta t$ with $\Delta t<<\gamma_b^{-1}$. At $t=\Delta t$, an additional motor protein can bind to the microtubule with probability $p=\gamma_b\Delta t$. If an additional plus motor binds or no additional motor protein binds to the microtubule, the cargo will continue to move to the same direction. If a minus motor binds to the microtubule, the cargo will stop moving. At every $t=m\Delta t$ instant, a motor protein attempts to bind to the microtubule with probability $p$. This process repeats {\em ad infinitum}. At a given instant, the majority motor proteins move the cargo towards their own moving direction. The velocity of a cargo can be determined as a function of the number of active plus motors $n_+$ and that of active minus motors $n_-$, which depends on the specific transport models. In one of the models, it is assumed that the minority motor proteins act as a load to the motion of a cargo driven by majority motor proteins. By balancing the forces acting on two kinds of motor proteins and equating their velocities, the velocity of a cargo can be determined.\cite{Lipowsky} In our model, we will assume that the motor coordination mediated by a {\em tug-of-war} mechanism forces the minority motor proteins to turn off and not to participate in moving the cargo. The cargo is solely driven by the majority motor proteins at constant velocity $v$.\cite{Wieschaus}

\begin{figure}
\epsfxsize=5.in \epsfysize=4.in \epsffile{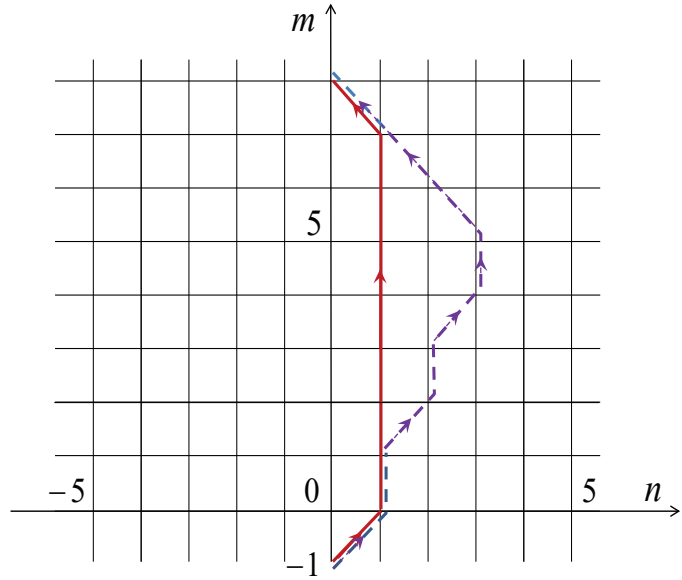}
\caption{(Color online) Two paths forming half-loops. The net number of motor proteins $n=n_{+}-n_-$ is plotted as a function of discrete time $m$. For these two paths, the majority motor proteins remain to be plus motors and hence a cargo moves unidirectionally to the plus-end of microtubule until it stops at $m=8$.}
\label{HalfLoop}
\end{figure}

\begin{figure}
\epsfxsize=4.in \epsfysize=4.in \epsffile{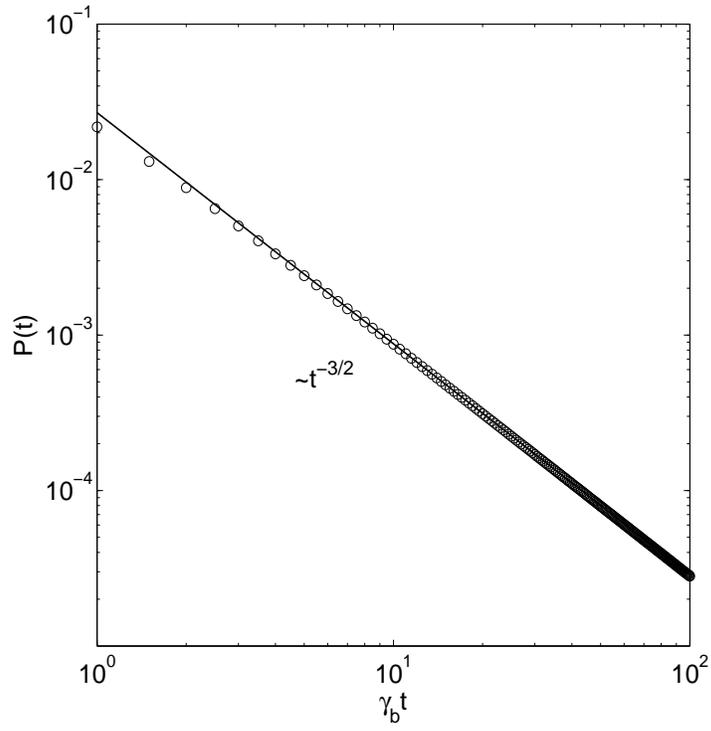}
\caption{The probability $P(t)$ for half-loop paths with run time $t$.  The circles represent the numerical result for $P(t)$ as a function of discrete time $\gamma_b t$ with $t=m\Delta t$. The solid line is a linear fit to the data showing the power-law behavior given by $P(t)\propto t^{-3/2}$.}
\label{LevyWalk}
\end{figure}

For the moment, we will neglect any correlation effect involved in binding process of a motor protein to microtubule, which we will assume to be a Markov process and will take $\gamma_u=0$. In Fig.~\ref{HalfLoop}, the net number of motor proteins $n=n_{+}-n_-$ is plotted as a function of discrete time $m\Delta t$. Two paths are drawn by (red)solid and (blue)dashed lines, where the values of $n$ remain to be positive for $m<8$. They form half-loop paths. One can notice that for the above two paths, a cargo will move unidirectionally to the plus-end until $m=8$ and then stop for the moment. Every distinct half-loop path with the identical ending time $t$ corresponds to the same unidirectional motion of a cargo with run time $t$. Now we want to calculate the probability $P(m)$ to form half-loops with run time $t=m\Delta t$. We will define the probability distribution function $u(n,m)$, which represents the probability for the net number of motors to be $n$ at $t=m\Delta t$.
We will restrict ourselves to the paths, which are localized to the positive side of $n$ with the initial condition of $u(n,0)=\delta_{n1}$. For the half-loop paths, the time evolution of $u(n,m)$ is governed by the following recurrence relations
\begin{eqnarray}
u(1,m+1)=r u(1,m)+p u(2,m)\nonumber \\
u(n,m+1)=p u(n-1,m)+r u(n,m)+p u(n+1,m)\,\,\,\, {\rm for}\,\,\,n\ge 2.\nonumber \\
\label{Evolution_Loop}
\end{eqnarray}
with $r=1-2p$.
We will define a function $S(m)=\sum_{n=1}^{m+1} u(n,m)$, which represents the probability for a cargo to remain localized to the positive side until $t=m\Delta t$. From the above recursion relations, one can obtain the following equation: $S(m+1)=-p u(1,m)+S(m)$. Hence the probability $P(m+1)$ for half-loop paths with the ending time $m+1$ is identical to the escape probability given by the following formula
\begin{equation}
P(m+1)=p u(1,m)=-\left(S(m+1)-S(m)\right).
\label{Prob_Loop}
\end{equation}
Using the recursion relations, we have numerically calculated the probability $P(t)$ for half-loop paths with run time $t=m\Delta t$.
In Fig.~\ref{LevyWalk}, the probability $P(t)$ is plotted as a function of $\gamma_b t$. The circles represent the numerical result for $P(t)$ and the solid line is a linear fit to the data demonstrating the power-law behavior given by $P(t)\propto t^{-3/2}$.
One can also calculate the probability $P_0(t)$ that no motor proteins bind to the microtubule until $t$ and hence the cargo pauses during $t$. It can be obtained by using the following formula: $P_0(t)=\lim_{m\rightarrow\infty} \left(1-2\gamma_b t/m\right)^m=e^{-2\gamma_b t}$. Hence it follows a Poisson distribution.

We now want to calculate the probability $P(t)$ analytically and also consider the effect of finite unbinding rate $\gamma_u$.
After time $t>\gamma_b^{-1}$ has elapsed, one can assume that smooth distribution of $u(n,m)$ has reached. One can derive the continuum diffusion equation with absorbing boundary condition at $x=0$ based on Eq.~\ref{Evolution_Loop}, which can be written by
\begin{equation}
{\partial u(x,t)\over \partial t}= \gamma_b {\partial^2 u(x,t)\over \partial x^2} \,\,\,\, {\rm with}\,\,\,\, u(0,t)=0,
\label{Continuum_Loop}
\end{equation}
where $x$ represents the continuum version of $n$. One can obtain the exact solution of Eq.~\ref{Continuum_Loop}, which will be briefly shown below.\cite{Chabdrasehkar} We will define the following basis function $\phi (k,x)=\sqrt{2/\pi}\sin kx$ with $k>0$, which automatically satisfy the absorbing boundary condition $\phi (k,0)=0$. One can express the function $u(x,t)$ as an integral over $k$ using the basis function $\phi (k,x)$ as follows: $u(x,t)=\int_0^\infty dk {\tilde u}(k,t) \phi(k,x)$.
By putting $u(x,t)$ into Eq.~\ref{Continuum_Loop}, one can obtain the following equation $d{\tilde u}/dt=-\gamma_b k^2{\tilde u}$, whose solution can be written by
\begin{equation}
{\tilde u}(k,t)={\tilde u}(k,0) e^{-\gamma_b k^2 t}.
\label{U_tilde}
\end{equation}
We will assume that a single plus motor has been bound to microtubule at $t=0$ and hence $u(x,0)=\delta(x-1)$.
Following the straightforward algebra, one can obtain $u(x,t)$ as follows
\begin{equation}
u(x,t)={1\over\sqrt{4\pi \gamma_b t}}\left[e^{-(x-1)^2/(4\gamma_b t)}-e^{-(x+1)^2/(4\gamma_b t)}\right].
\label{Uxt}
\end{equation}
Hence $u(x,t)$ can be viewed as a difference between normal distribution centered at $x=1$ and that centered at $x=-1$.

The effect of finite unbinding rate $\gamma_u$ of bound motor proteins can be taken into account as follows. Starting from $n=1$ at $t=0$, binding of an additional motor protein is attempted every $\Delta t$ seconds. At $t=m\Delta t$, the net effective number $n_{\rm eff}$ of bound motor proteins can be written by $n_{\rm eff}=e^{-\gamma_u m\Delta t}+\sum_{l=1}^{m}e^{-\gamma_u (m-l)\Delta t}\sigma_l$. Here $\sigma_l=0, \pm 1$ correspond to the cases of no motor protein$(0)$, a plus motor$(+1)$, and a minus motor$(-1)$ respectively. One can show that the mean value of the net effective number $\left\langle n_{\rm eff}\right\rangle$ at $t=m\Delta t$ is given by ${\bar n}(t)=e^{-\gamma_u t}$ and the variance is given by $\Delta n_{\rm eff}^2=\left\langle n_{\rm eff}^2\right\rangle-\left\langle n_{\rm eff}\right\rangle^2=\gamma_b\gamma_u^{-1}(1-e^{-2\gamma_u t})$. For relatively short time $t<\gamma_u^{-1}$, $\Delta n_{\rm eff}^2\cong 2\gamma_b t$ and hence one can read off the diffusion constant to be $\gamma_b$. Since $n_{\rm eff}$ is a linear combination of random variables $\sigma_l$ with deterministic coefficients, it will follow gaussian distribution with the modified diffusion constant and the modified mean value. Hence $t$ is replaced with the effective time $\Gamma(t)=(1-e^{-2\gamma_u t})/(2\gamma_u)$ and the mean value is replaced with ${\bar n}(t)$.
With the inclusion of finite unbinding rate $\gamma_u$, one can finally obtain $u(x,t)$ as follows
\begin{equation}
u(x,t)={1\over\sqrt{4\pi \gamma_b \Gamma(t)}}\left[e^{-(x-{\bar n}(t))^2/4\gamma_b\Gamma(t)}-e^{-(x+{\bar n}(t))^2/4\gamma_b\Gamma(t)}\right].
\label{unbinding}
\end{equation}

We are now ready to calculate the probability per unit time $P(t)$ to have half-loops with run time $t$. We will define the function $S(t)$ as follows
\begin{equation}
S(t)=\int_0^\infty dx u(x,t)={\rm erf}\left({\bar n}(t)/\sqrt{4\gamma_b\Gamma(t)}\right).
\label{Prob_Loop_continuum}
\end{equation}
The probability per unit time $P(t)$ can be calculated by using the following formula $P(t)=-\partial S(t)/\partial t$, which is given by
\begin{equation}
P(t)=\frac {{\bar n}(t)}{\sqrt{4\pi\gamma_b\Gamma^3(t)}}\,e^{-{\bar n}^2(t)/(4\gamma_b\Gamma(t))}.
\end{equation}
For short time $t<\gamma_b^{-1}$, since an additional minus motor should bind to microtubule to form a loop, the probability $P(t)$ starts to increase from $0$ at $t=0$ with time $t$.
For the intermediate time $\gamma_b^{-1}<t<\gamma_u^{-1}$, the probability $P(t)$ can be written by $P(t)\cong t^{-3/2}/\sqrt{4\pi\gamma_b}$. One can clearly notice that the power-law behavior observed in our numerical result shows an excellent agreement with the analytical one. For relatively long time $t>\gamma_u^{-1}$, $P(t)\cong e^{-\gamma_u t}\,/\sqrt{\pi\gamma_b/2\gamma_u^3}$.

\begin{figure}
\epsfxsize=4.in \epsfysize=4.in \epsffile{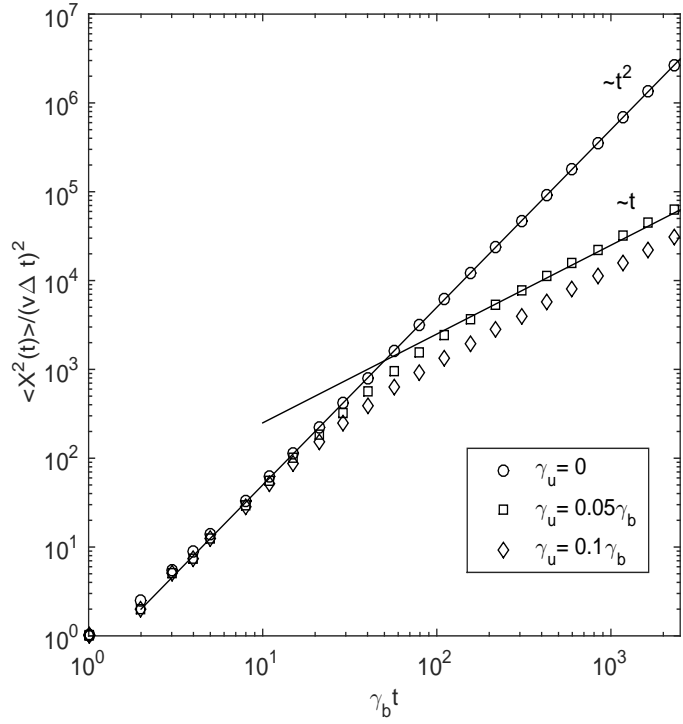}
\caption{Mean squared displacement $\langle X^2(t)\rangle$ of a cargo for $\gamma_u/\gamma_b=0, 0.05, 0.1$. $\langle X^2(t)\rangle/(v\Delta t)^2$ is plotted as a function of $\gamma_b t$. For $t<\gamma_u^{-1}$, it follows super-diffusive behavior $\left\langle X^2(t)\right\rangle\propto t^2$, which changes to normal diffusion $\left\langle X^2(t)\right\rangle\propto t$ for relatively long time $t>\gamma_u^{-1}$.
}
\label{MSDofCargoGamma_u}
\end{figure}

We now study the dynamical properties of a cargo transport. 
The run length $L(t)$ of a cargo corresponding to a half loop with run time $t$ is simply given by $L(t)=vt$.
Starting from the origin at $t=0$, the position $X(t)$ of a cargo at time $t=m\Delta t$ can be written by
\begin{equation}
X(t)/({v\Delta t})=\sum_{l=1}^m sign\left(\sum_{n=1}^le^{-\gamma_u(l-n)\Delta t}\sigma_n\right).
\label{Analytic_MSD1}
\end{equation}
We numerically calculate the mean squared displacement(MSD) $\left\langle X^2(t)\right\rangle$ with finite unbinding rate $\gamma_u$.
By taking an average over $10^5$ independent trajectories following the recursion relations, we have obtained the MSD of a cargo as a function of discrete time $t=m\Delta t$ for three different values of $\gamma_u/\gamma_b=0, 0.05, 0.1$.
In Fig.~\ref{MSDofCargoGamma_u}, the MSD is plotted as a function of $\gamma_b t$.
The circles represent the MSD for $\gamma_u=0$. The solid curve is a linear fit to the data, which demonstrates super-diffusive behavior $\langle X^2(t)\rangle\propto t^2$. The squares and diamonds represent the MSD for $\gamma_u/\gamma_b=0.05, 0.1$ respectively.
For $t<\gamma_u^{-1}$, it follows super-diffusive behavior $\left\langle X^2(t)\right\rangle\propto t^2$, which changes to normal diffusion $\left\langle X^2(t)\right\rangle\propto t$ for relatively long time $t>\gamma_u^{-1}$.

Finally, we want to investigate the important correlation effect involved in the binding process of motor protein to microtubule. We will propose that the motor coordination mediated by a {\em tug-of-war} mechanism not only determines the moving direction and velocity of a cargo but also influences the subsequent binding of an additional motor in the following way. When the cargo is moving towards the plus-end of microtubule at certain instant, the binding rate of an additional plus motor increases, while the binding rate of an additional minus motor decreases. We will call this case as `positive correlation' and the opposite as `negative correlation'.
The binding rates of plus and minus motors are given by $\gamma_{b+}=\gamma_b+\gamma_c, \gamma_{b-}=\gamma_b-\gamma_c$ respectively. For positive correlation $\gamma_c>0$, binding of an additional majority motor protein will be more favored than that of an additional minority motor protein. For negative correlation $\gamma_c<0$, binding of an additional minority motor protein will be more favored than that of an additional majority motor protein.
For the sake of clarity, we will take $\gamma_u=0$.
We now calculate the probability per unit time $P(t)$ analytically. In spite of the correlation effect, one can obtain the exact solution. For the half-loops localized to the positive side of $n$, the cargo is always moving to the plus-end until it stops. Hence we simply need to put $\gamma_{b+}$ for binding of a plus-motor and $\gamma_{b-}$ for binding of a minus-motor. One can obtain the modified continuum diffusion equation with absorbing boundary condition at $x=0$, which can be written by
\begin{equation}
{\partial u_c(x,t)\over \partial t}= \gamma_b {\partial^2 u_c(x,t)\over \partial x^2}-2\gamma_c{\partial u_c(x,t)\over \partial x} \,\,\,\, {\rm with}\,\,\,\, u_c(0,t)=0.
\label{Correlation}
\end{equation}
One can show that the solution of Eq.~\ref{Correlation} can be written by $u_c(x,t)=e^{-\left(\gamma_c^2/\gamma_b\right)t} e^{\left(\gamma_c/\gamma_b\right)(x-1)}u(x,t)$, where
$u(x,t)$ is given by Eq.~\ref{Uxt}. The probability per unit time $P(t)=-\partial S(t)/\partial t$ can be written by
\begin{equation}
P(t)=\frac {1}{\sqrt{4\pi\gamma_b}}\,t^{-3/2}\,e^{-\gamma_c/\gamma_b}\,e^{-1/(4\gamma_b t)-(\gamma_c^2/\gamma_b)t}.
\label{P_correlation}
\end{equation}
For $\gamma_b^{-1}<t<\gamma_b/\gamma_c^2$, $P(t)\cong t^{-3/2}\,e^{-\gamma_c/\gamma_b}/\sqrt{4\pi\gamma_b}$. It is quite remarkable to notice that L{\'e}vy walk behavior persists robustly against correlations only adding an effective cutoff time $\gamma_b/\gamma_c^2$.
Since plus(minus) motors are preferred to be bound for positive(negative) correlation, one might expect that for a given run time $t$, the positive(negative) correlations will tend to increase(decrease) the probability $P(t)$ in comparison to that without correlations.
However, contrary to our intuition, $P(t)$ decreases exponentially for relatively long time $t>\gamma_b/\gamma_c^2$ regardless of the sign of correlations. This can be understood as follows. For the half-loops with a fixed run time $t$, the number of plus motors and that of minus motors should be the same and hence the probability to generate half loops with $n_{+}=n_{-}=N$ will always get reduced by a factor of $(\gamma_{b+}\gamma_{b-}/\gamma_b^2)^N=[1-(\gamma_c/\gamma_b)^2]^N$.

In summary, we have studied the cargo transport along microtubule by motor coordination mediated by a {\em tug-of-war} mechanism. We have shown that the probability $P(t)$ for unidirectional run time $t$ of a cargo exhibits L{\'e}vy walk behavior $P(t)\propto t^{-{3/2}} e^{-\gamma_u t}$ with a finite cutoff time $\gamma_u^{-1}$. By considering the correlation effect in binding process of a motor protein to microtubule, we have demonstrated that L{\'e}vy walk behavior of $P(t)\propto t^{-3/2}$ remains to hold only with an introduction of effective cutoff time $\gamma_b/\gamma_c^2$. Although the behavior of $P(t)$ as a function of run length $L(t)$ may vary depending on the specific transport models used, we conjecture that L{\'e}vy walk statistics of $P(t)\propto t^{-{3/2}}$ as a function of run time $t$ can be relatively robust. Our findings can be generally applicable to various phenomena in which the underlying mechanism is governed by majority rule following stochastic voting.

K.~M. wishes to acknowledge the financial support by Basic Science Research Program through the National Research Foundation of Korea(NRF) funded by the Ministry of Education, Science and Technology(NRF-2016R1D1A1B01013756).

\newpage

\clearpage

\clearpage

\clearpage

\end{document}